\documentclass[]{article}
\usepackage{graphicx}

\begin{document}

\centerline{\Large \bf Does Sex Induce a Phase Transition ?}\medskip

\bigskip

\centerline{P.M.C. de Oliveira$^{1,2}$, S. Moss de Oliveira$^{1,2}$, D. 
Stauffer$^{2,3}$, S. Cebrat$^4$ and A. P\c{e}kalski$^5$}

\bigskip\noindent
$^1$ Instituto de F\'{\i}sica, Universidade Federal Fluminense; Av. 
Litor\^{a}nea s/n, Boa Viagem, Niter\'{o}i 24210-340, RJ, Brazil

\medskip\noindent
$^2$ Laboratoire PMMH, \'Ecole Sup\'erieure de Physique et de Chimie 
Industrielles, 10 rue Vauquelin, F-75231 Paris, France

\medskip\noindent
$^3$ Institute for Theoretical Physics, Cologne University, D-50923 K\"oln, 
Euroland

\medskip\noindent
$^4$ Department of Genomics, Wroc{\l}aw University, ul. Przybyszewskiego 63/77, 
51-148 Wroc{\l}aw, Poland

\medskip\noindent
$^5$ Institute of Theoretical Physics, Wroc{\l}aw University, pl. Maxa Borna 9, 
50-204 Wroc{\l}aw, Poland

\bigskip\noindent
e-mail address: pmco@if.uff.br

\bigskip

\begin{abstract}

	We discovered a dynamic phase transition induced by sexual 
reproduction. The dynamics is a pure Darwinian rule applied to diploid 
bit-strings with both fundamental ingredients to drive Darwin's evolution: 1) 
random mutations and crossings which act in the sense of increasing the entropy 
(or diversity); and 2) selection which acts in the opposite sense by limiting 
the entropy explosion. Selection wins this competition if mutations performed 
at birth are few enough, and thus the wild genotype dominates the steady-state 
population. By slowly increasing the average number $\, m\, $ of mutations, 
however, the population suddenly undergoes a mutational degradation precisely 
at a transition point $\, m_{\rm c}\, $. Above this point, the ``bad'' alleles 
(represented by 1-bits) spread over the genetic pool of the population, 
overcoming the selection pressure. Individuals become selectively alike, and 
evolution stops. Only below this point, $\, m < m_{\rm c}\,$, evolutionary life 
is possible.

	The finite-size-scaling behaviour of this transition is exhibited for 
large enough ``chromosome'' lengths $\, L\, $, through lengthy computer 
simulations. One important and surprising observation is the $L$-independence 
of the transition curves, for large $\, L\, $. They are also independent on the 
population size. Another is that $\, m_{\rm c}\,$ is near unity, i.e. life 
cannot be stable with much more than one mutation per diploid genome, 
independent of the chromosome length, in agreement with reality. One possible 
consequence is that an eventual evolutionary jump towards larger $\, L\, $ 
enabling the storage of more genetic information would demand an improved DNA 
copying machinery in order to keep the same total number of mutations per 
offspring.

\end{abstract}

\newpage
 \section{Introduction}

	The theoretical question posed in this work concerns the length-scaling 
properties of chromosomes. Let's call $\, L\, $ the chromosome length, an 
integer number measuring the number of coding units along the chain, which for 
simplicity we consider as a bit-string: 0-bits represent the wild alleles, 
whereas 1-bits correspond to harmful mutations, the ``bad'' alleles. The larger 
this length $\, L\, $ is, the larger is the space to store more genetic 
information. Therefore, in principle, evolution should lead to species with 
larger and larger chromosomes, of course with the same value of $\, L\, $ for 
all individuals belonging to the same species.

	Consider first a simple case of haploid individuals which reproduce 
through cloning. The chromosome of each newborn is copied from an already alive 
individual, taken at random, plus an average fixed number $\, m\, $ of point 
mutations. Being an average over all newborns, this number $\, m\, $ is not 
necessarily an integer, it can be tuned in a continuously way as explained 
later. One point mutation means a 0-bit in the parent's chromosome which is 
flipped into a 1-bit in the offspring's, or vice-versa. The position where this 
mutation is performed is random. The wild genotype corresponds to a bit-string 
where all bits are set to zero. A mutation in the sense $\, 0 \to 1\, $ makes 
the offspring farther to the wild genotype than its parent, another in the 
reverse sense makes it closer. A fixed birth rate $\, b\, $ defines the 
probability of each individual to produce an offspring each new time step.

	Let's ignore any kind of correlation along the chromosome, i.e. the 
fitness of individual $\, i\, $ depends only on a single phenotype defined here 
as $\, N_i\, $, the total number of 1-bits in its genome. One individual with 
phenotype $\, N+1\, $ is at a disadvantage, when compared to another individual 
with phenotype $\, N\, $. The disadvantage here corresponds to a smaller 
survival chance: the probability to survive a new time step is smaller for the 
former individual by a factor of $\, x\, $, when compared to the latter, where 
$\, x\, $ is a number strictly smaller than 1. Therefore, the survival 
probability for different individuals decrease for increasing $\, N\, $. This 
number $\, x\, $ measures the overall selection pressure, and can be tuned in 
order to keep the population size constant, i.e. to keep the death rate equal 
to the birth rate $\, b\, $. After evolving for many generations the 
distribution of phenotypes stabilizes. In order to keep the wild genotype (the 
only for which the phenotype is $\, N = 0\, $) inside this equilibrium 
distribution, the number of mutations $\, m\, $ cannot be too high.

	Let's now compare different chromosome lengths. One can follow a simple
and intuitive reasoning:

\bigskip

	1) the length $\, L\, $ is increased;

\smallskip

	2) the same ratio $\, m/L\, $ is kept;

\smallskip

	3) after many generations, the steady-state population presents the 
same distribution of phenotypes {\sl versus} $\, N/L\, $, independent of the 
(large enough) chromosome length.

\bigskip

	This expected behaviour is exactly what is obtained by simulating this
simple haploid, asexual model on a computer \cite{book,pre}. Fig.1 shows an 
example of this behaviour \cite{jpc}.

\begin{figure}[!hbt]

\begin{center}
 \includegraphics[angle=-90,scale=0.5]{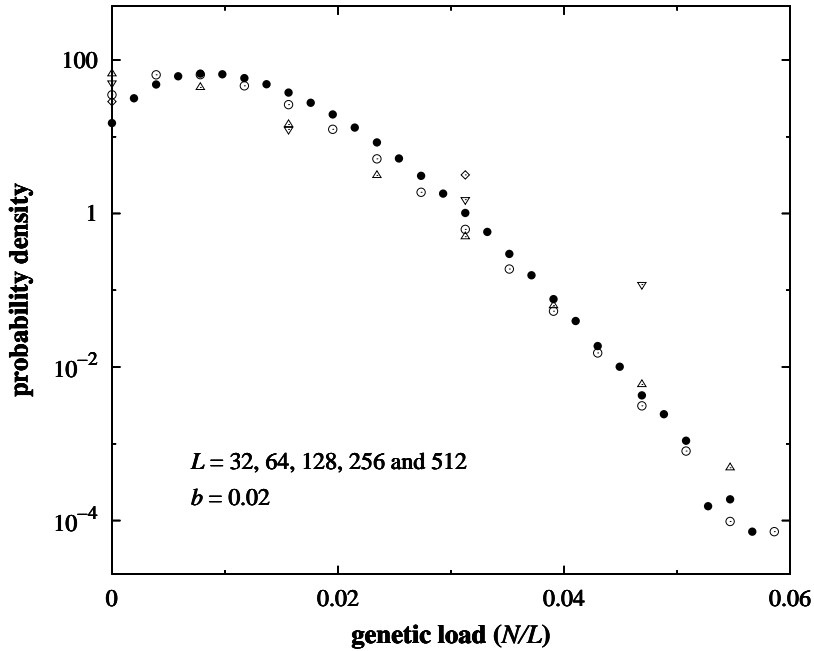}
\end{center}

\caption{Collapsed distributions of the individual genetic loads $\, N/L\, $, 
for haploid, asexual reproducing populations with different chromosome lengths. 
The probability density plotted along the vertical axis is proportional to the 
number of individuals sharing the same $\, N\, $. The full circles correspond 
to the largest length $\, L = 512\, $. The mutation rate $\, m/L = 1/320 
\approx 0.003\, $ is the same for all lengths, as well as the population size 
$\, P = 10000\, $.}

\label{fig1}
\end{figure}

	The above-mentioned item 2) deserves an important remark: the genetic 
storing media (the bit-strings) are one-dimensional objects. Therefore, the 
average number $\, m\, $ of mutations should be scaled proportionally to $\, 
L\, $. As a result, the whole genetic distribution curve and consequently both 
its average $\, \langle N\rangle\, $ and its width $\, \langle\Delta N\rangle\, 
$ also scale proportionally to $\, L\, $ (note the collapsed distribution 
curves in Fig.1 plotted {\sl versus} $\, N/L\, $, not $\, N\, $).

	The reasoning and the corresponding simulational results do not cause 
any surprise. The purpose of this work is to study a similar reasoning for 
sexual, diploid reproduction. Let's pose the first question.

\bigskip

	Should the same ratio $\, m/L \, $ be kept for increasing chromosome 
lengths?

\bigskip

\noindent The answer to this simple question is not so simple. Intuition can 
betray who thinks about it. Sex deals with half the genetic information 
inherited from each parent, a nonlinear behaviour which requires prudence to 
avoid false conclusions. Moreover, a crossing-over performed with homologous 
chromosomes within each parent's genome indicates that now the genetic 
information is no longer stored along strictly one-dimensional objects: we 
should not trust on the linear reasoning leading to the fixed ratio $\, m/L\, 
$. Dominance and recessiveness are further sources of doubts. In order to 
answer this and many other related questions, we present in the next sections 
the results obtained from computer simulations of a population dynamics. 
Compared to reality, the model is simplified in order to retain only the 
fundamental features of sexual, diploid reproduction. It is based on a pure 
Darwinian evolutionary rule with two basic ingredients: random mutations which 
tends to increase the entropy (or diversity); and natural selection which acts 
on the opposite sense by removing from the population many of these mutations 
and, consequently, preventing entropy explosion.

\newpage
 \section{Conceptual remarks}

	Fig.2 shows a computer-simulated result of a model described later on. 
For the moment, some general informations are enough. First, it considers a 
sexually reproducing population, with individual genomes subjected to random 
mutations as well as crossing-over during reproduction, under a selective 
environment. Homologous chromosomes are represented by double, diploid 
bit-strings consisting of 0-bits (the wild type) and 1-bits (the ``bad'' 
allele). The quantity we consider for selection is the total number $\, N\, $ 
of loci containing at least one ``bad'' allele; the larger $\, N\, $ is, the 
smaller is the individual's fitness. Data in Fig.2 correspond to the average 
over 10 independent stable populations (after many enough generations). All 
other data presented in this paper also correspond to the average over 10 
independent populations. Error bars were determided from fluctuations between 
these 10 samples.

\begin{figure}[!hbt]

\begin{center}
 \includegraphics[angle=-90,scale=0.5]{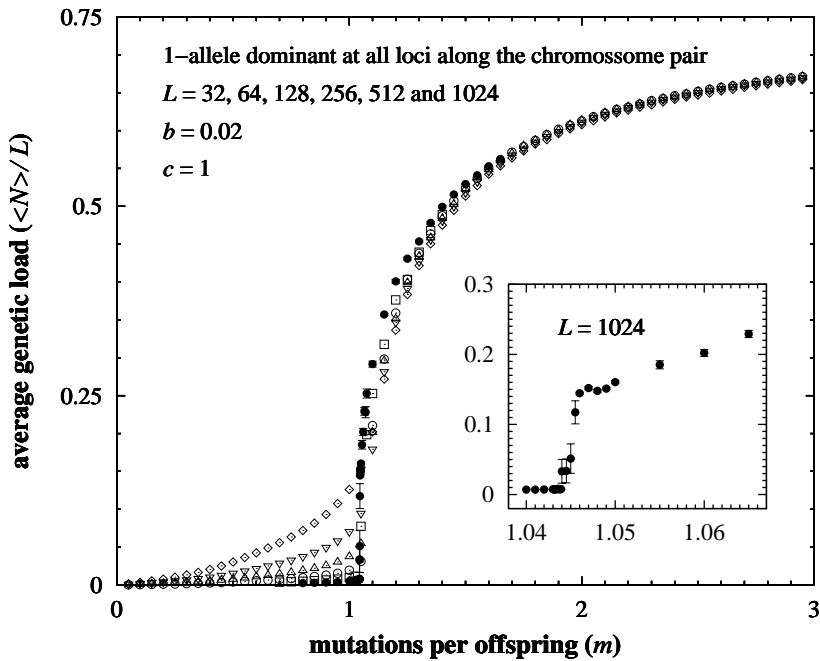}
\end{center}

\caption{Phase transition for a population of $\, 10000\, $ individuals. For 
few enough mutations at birth, on the left side, life is possible. Beyond the 
sharply defined point $\, m_c = 1.0439\, $, on the right side, the population 
displays a mutational explosion where all individuals carry a number of ``bad'' 
alleles proportional to the chromosome length $\, L\, $. For the largest 
length, $\, L = 1024\, $, the inset blows-up the transition region.}

\label{fig2}
\end{figure}

	Fig.2 shows plots for different chromosome lengths, and one verifies at 
the left side the corresponding curves approaching the horizontal axis for 
larger and larger $L\, $. The average genetic load $\, \langle N\rangle/L\, $, 
where the symbol $\, \langle\dots\rangle\, $ means population average, vanishes 
for large enough chromosome lengths along {\sl all} this phase, $\, m < m_{\rm 
c}\, $.

	Instead, on the right side of Fig.2, the curves go {\sl up}.  By 
increasing the chromosome length, they converge to the universal curve 
displayed by full-circles ($\, L = 1024\, $). The average genetic load is no 
longer null, because $\, \langle N\rangle\, $ becomes proportional to $\, L\, 
$: life through Darwinian selection becomes impossible, as explained below.

\begin{figure}[!hbt]

\begin{center}
 \includegraphics[angle=-90,scale=0.5]{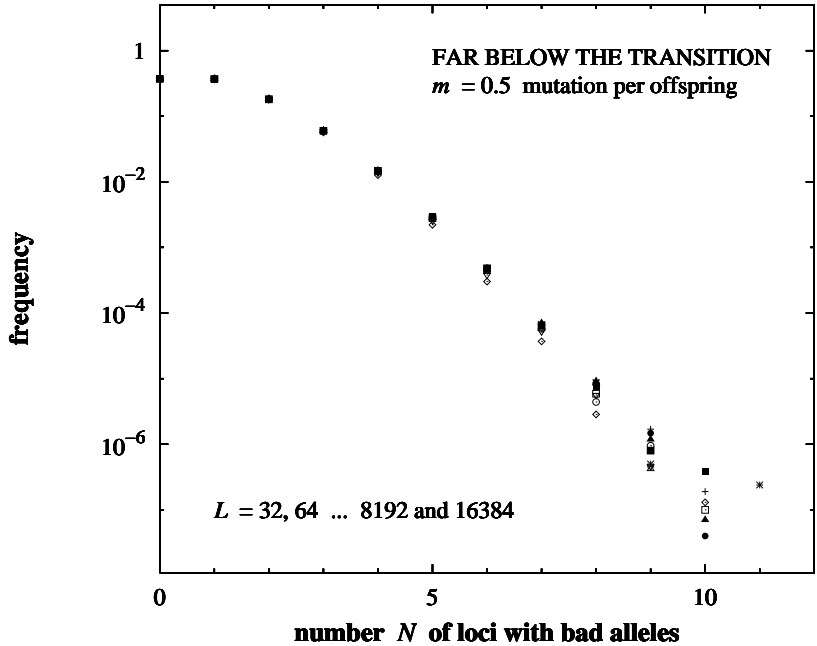}
\end{center}

\caption{Collapsed distributions of ``bad'' alleles among the population, for 
different chromosome lengths. Note that $\, N\, $, displayed along the 
horizontal axis, {\sl is not} divided by $\, L\, $, contrary to the asexual 
case, Fig.1. In all cases, the number $\, m\, $ of mutations per offspring 
performed at birth is fixed far below the transition value observed in Fig.2, 
i.e. $\, m = 0.5\, $, again not divided by $\, L\, $. Would we fix the same 
mutation rate $\, m/L\, $, instead of $\, m\, $, the plots would no longer 
collapse onto each other. Moreover, in this case, for large $\, L\, $ the 
curves would undergo a run-away to the right as soon as the value of $\, m\, $ 
surpasses the critical point $\, m_{\rm c} = 1.0439\, $ of Fig.2, as explained 
soon. The parameter controlling the phase transition is the number of mutations 
$\, m\, $, not the mutation rate $\, m/L\, $.}

\label{fig3}
\end{figure}

	Fig.3 shows the distributions of homologous loci containing ``bad'' 
alleles. For larger and larger chromosome lengths, all curves collapse into a 
single one. The data are collected below the transition, deeply inside the 
ordered phase on the left side of Fig.2, $\, m = 0.5\, $. The typical number 
$\, \langle N\rangle\, $ of loci containing the ``bad'' allele ($\, \langle 
N\rangle < 5\, $ in Fig.3) remains the same in spite of the increasing 
chromosome lengths. That is why the curves go {\sl down} on the left side of 
Fig.2 as

$$\frac{\langle N\rangle}{L}\, \propto\, L^{-1}$$

\noindent where the symbol $\, \propto\, $ represents proportionality. The 
exponent $\, -1\, $ governs the finite-size-scaling of the genetic load $\, 
\langle N\rangle/L\, $, and is our first important result.

\begin{figure}[!hbt]

\begin{center}
 \includegraphics[angle=-90,scale=0.5]{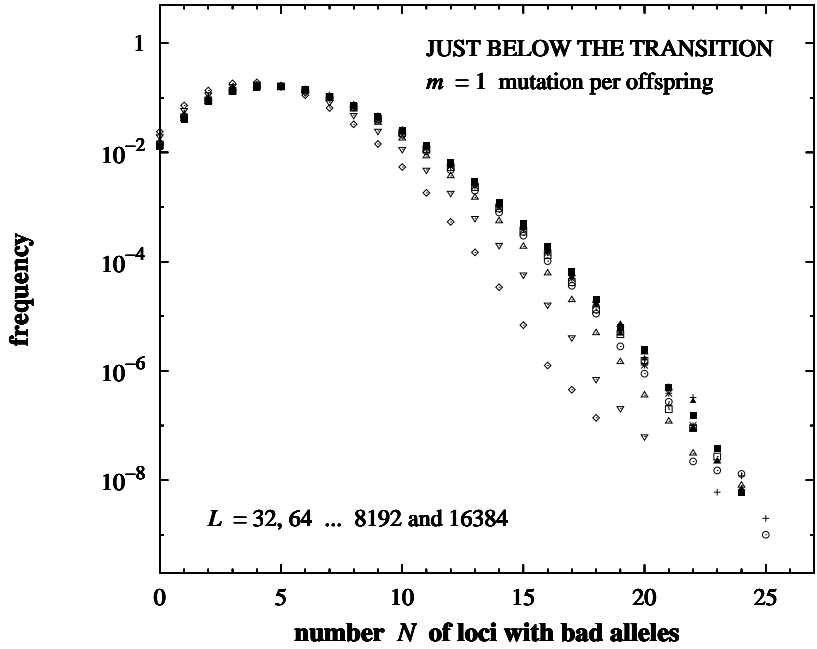}
\end{center}

\caption{Distribution of ``bad'' alleles for different chromosome lengths, as 
in Fig.3. Now, the number $\, m\, $ of mutations per offspring performed at 
birth is fixed just below the transition displayed in Fig.2, $\, m = 1\, $. For 
large enough $\, L\, $ the curves also collapse onto each other.}

\label{fig4}
\end{figure}

	Fig.4 shows again the $\, N$-distribution for the same transition 
displayed in Fig.2. Now, the average number of mutations performed at birth is 
$\, m = 1\, $ for all chromosome lengths, very near but still below the 
transition point $\, m_{\rm c} = 1.0439\, $ of Fig.2. The typical number $\, 
\langle N\rangle \approx 5\, $ of loci containing the ``bad'' allele is larger 
now, when compared with Fig.3. However, it also remains the same for increasing 
chromosome lengths. Note also that the ``optimum'' configuration $\, N = 0\, $ 
is still present, although with a small frequency.

\begin{figure}[!hbt]

\begin{center}
 \includegraphics[angle=-90,scale=0.5]{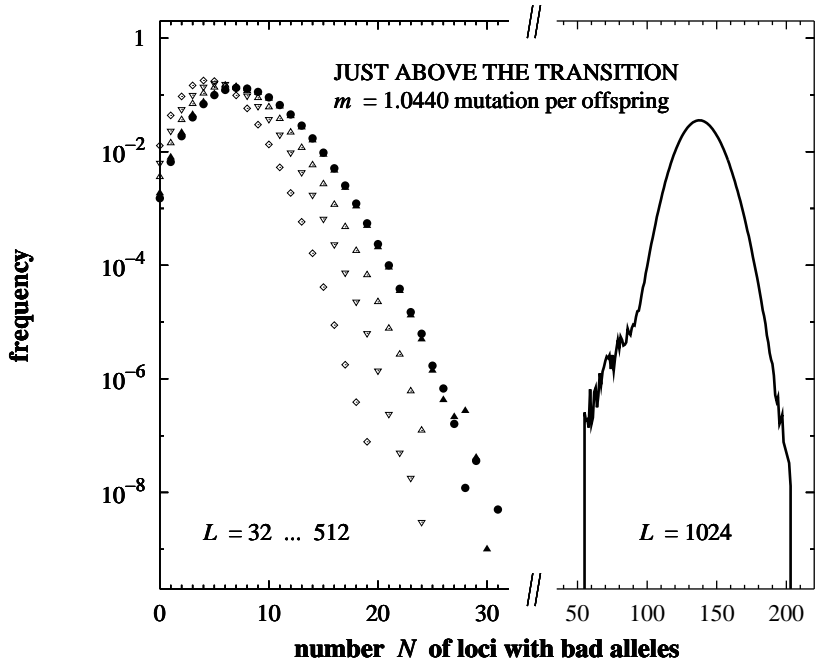}
\end{center}

\caption{The same as the previous two figures, now with $\, m = 1.0440\, $, 
just above the transition displayed in Fig.2. For large enough chromosome 
lengths, the distribution runs away from the wild genotype represented here by 
$\, N = 0\, $. The rightmost continuous line obtained for $\, L = 1024\, $ 
shows this behaviour, the whole distribution being confined in between the two 
vertical walls.}

\label{fig5}
\end{figure}

	At each new time step, a fraction $\, b\, $ of new individuals are 
included into the population. We adopted $\, b = 2\%\, $. Their genomes are 
taken from random parents, with mutations. Since the number of 0-bits among the 
population is much larger than that of 1-bits, these mutations occur more 
likely in the sense $\, 0 \to 1\, $ (``bad'' mutations) than in the reverse 
one, as in Nature. This asymmetry tends to shift the curves like Fig.4 to the 
right, increasing its rightmost parts. Selection, which eliminates the same 
fraction $\, b\, $ of individuals from the population, within the same time 
step, tends to shift the curves back to the left, in a compensatory movement. 
The figures show the steady-state situation, where the distribution remains the 
same after both movements were performed, i.e. after each computer time step 
with deaths and births. These two opposed movements, however, come from 
different ingredients of the Darwinian paradigm: the first (shifting the curve 
to the right) from random mutations which we control through the parameter $\, 
m\, $; the second (back to the left) from the selection pressure which is 
always the same, since we keep the death rate $\, b\, $ constant. Therefore, by 
further increasing $\, m\, $, this balance which keeps the wild genotype alive 
will become impossible.

	Fig.5 corresponds to $\, m = 1.0440\, $, just beyond the transition. 
Indeed, the curves falsely seem to obey the same kind of convergence displayed 
in Fig.3 or 4, up to $\, L = 512\, $. Suddenly, however, for $\, L = 1024\, $ 
the distribution escapes towards a finite-density $\, \langle N\rangle /L\, $ 
of loci containing the ``bad'' allele, shown by the rightmost curve where $\, 
\langle N\rangle \approx 140\, $ (note the cut on the horizontal axis). The 
distribution curve runs away from the wild genomic form $\, N = 0\, $, which 
becomes extinct. This is the same phenomenon which occurs in Eigen-type models 
\cite{eigen}, sometimes called the ``error catastrophe''. Now, the typical 
number $\, \langle N\rangle\, $ of loci containing ``bad'' alleles grows 
proportionally to $\, L\, $: would we plot the distribution for $\, L = 2048$, 
the corresponding bell-shaped curve would be positioned around $\, \langle 
N\rangle \approx 280$, far to the right, not visible in Fig.5; for $\, L = 
4096\, $ it would fall around $\, \langle N\rangle \approx 560\, $, far yet to 
the right, and so on. We cannot show all these curves on the same plot: even 
for $\, L = 1024\, $ we were forced to perform the artificial cut on the 
horizontal axis.

\begin{figure}[!hbt]

\begin{center}
 \includegraphics[angle=-90,scale=0.5]{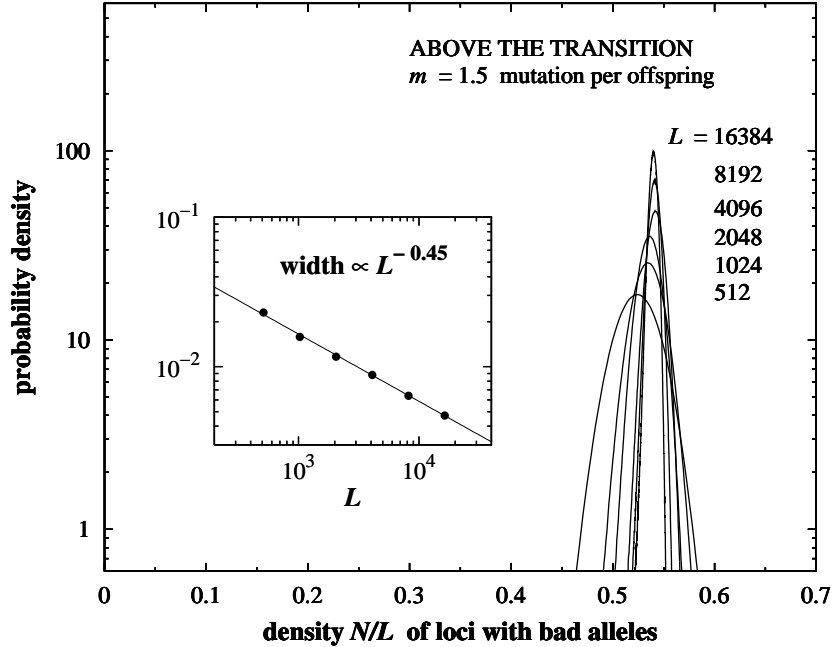}
\end{center}

\caption{Distributions for large enough chromosome lengths, above the 
transition. Now, they are plotted against the density $\, N/L\, $, as in the 
asexual case of Fig.1. However, contrary to the asexual case, the number $\, 
m\, $ of mutations performed at birth is kept fixed for different $\, L\, $, 
instead of the mutation rate $\, m/L\, $ kept fixed in Fig.1. The inset shows 
the corresponding $\, L$-dependence of the widths.}
 
\label{fig6}
\end{figure}

	In order to see the distribution curves for different chromosome 
lengths above the transition, we therefore replace (on the horizontal axis) the 
number $\, N\, $ of loci containing the ``bad'' allele by its {\sl density} $\, 
N/L\, $ along the genome. Fig.6 shows the result for $\, m = 1.5\, $.

	The widths $\, \Delta N/L\, $ of these distributions shrink for larger 
and larger chromosome lengths. Therefore, for large enough values of $\, L\, $, 
all individuals share the same genetic load $\, \langle N\rangle/L\, $, within 
negligible fluctuations. Individuals no longer show different selectivities 
when compared to each other, all individuals become alike in what concerns the 
selection pressure. Darwinian evolution cannot proceed for $\, m > m_{\rm c}\, 
$. This side of the transition is the non-evolutionary phase, and corresponds 
to population extinction as we shall see in next section.

	In short, the transition point $\, m_{\rm c}\, $ separates two phases. 
Region $\, m < m_{\rm c}\, $ represents the evolutionary phase where the 
typical number $\, \langle N\rangle\, $ of ``bad'' alleles remains the same for 
increasing chromosome lengths: the average genetic load $\, \langle 
N\rangle/L\, $ vanishes. The other phase, $\, m > m_{\rm c}\, $, is 
non-evolutionary and behaves differently: $\, \langle N\rangle\, $ increases 
proportionally to $\, L\, $, the genetic load no longer vanishes, and the 
genetic pool no longer includes the wild genotype $\, N = 0\, $. Note again 
that the transition occuring at $\, m_{\rm c}\, $ is controlled by the number 
of mutations $\, m\, $ (per genome), not by the mutation rate $\, m/L\, $ (per 
genome unit) , and consequently the transition point $\, m_{\rm c}\, $ remains 
the same, independently of how large is $\, L\, $.

	A dynamic phase transition corresponds to the competition of different 
possible attractors to which the steady-state population converges after many 
generations. In our case, one attractor is characterised by the presence of the 
wild genotype $\, N = 0\, $, which is preserved only in the ordered phase $\, m 
< m_{\rm c}\, $. We may call this evolutionary phase ``ordered'' by analogy 
with Physics where order-disorder transitions of this kind are ubiquitous, and 
also because the whole steady-state population remains ``orderly'' similar to 
the wild genotype, everybody with a vanishingly small fraction of ``bad'' 
alleles. On the other phase $\, m > m_{\rm c}\, $, the population genetic pool 
melts into a disordered situation characterised by the absence of the wild 
genotype $\, N = 0\, $, everybody presenting a non-vanishing fraction of 
``bad'' alleles. The selection mechanism is no longer able to contain the 
entropy explosion driven by too many random mutations at birth. The {\sl 
dis-}order parameter $\, \langle N\rangle/L\, $ characterises the transition, 
being non-null only at the disordered phase.

	A last comment concerning asexual reproduction \cite{jpc}. The same 
run-away shown for instance in Fig.5 also occurs in Fig.1. However, it does not 
correspond to a phase transition, because it can be avoided by increasing the 
population size. Genuine phase transitions require the so-called thermodynamic 
limit, where the size of the system under study goes to infinity. In practical 
terms, provided the population sizes are large enough, the collapsed curves in 
Fig.1 remain the same for larger and larger chromosome lengths: this behaviour 
characterises the absence of phase transitions. On the other hand, for our 
sexual case shown in Fig. 2, the transition point $\, m_{\rm c} = 1.0439\, $ 
does not move for different population sizes, which characterises the true 
existence of a phase transition.

\newpage
\section{The model}

	For the reader's convenience, the important conceptual results 
concerning this work are already discussed in both previous sections. The 
current one treats the implementation of the model on computers and its 
details. Other further results are in the next sections.

	The population size is artificially kept constant with $\, P\, $ 
individuals, by killing a fraction $\, b\, $ of them per time step and 
restoring the same fraction with newborns which are offspring produced by the 
survivors. The set of $\, P\, $ individuals is considered a random sample 
picked from a much larger population which can fluctuate in size, according to 
the selective dynamics. This is particularly important in case of extinction 
which occurs in the non-evolutionary phase. We verify that $\, P = 10^3\, $ or 
$\, 10^4\, $ is large enough to make the statistical fluctuations 
satisfactorily small for all quantities we have measured from our simulations. 
This $\, P\, $ is also large enough to avoid inbreeding depression 
\cite{sousa,bonkowsha}. We adopted $\, P = 10^4\, $ and $\, b = 0.02\, $ ($\, 
2\%\, $). The precise value of this fraction $\, b\, $ is not important, 
provided it is small enough, because it corresponds only to the rate according 
to which successive snapshots of the current population are taken, i.e. the 
``movie's speed''. We have also tested $\, b = 0.01\, $ and $\, 0.03\, $ in 
some cases, with the same results.

	The genetic information of each individual is kept on the computer 
memory in two parallel bit-strings with $\, L \,$ bits each. We tested $\, L = 
32\, $, 64, 128 $\dots$ 16384. We also keep on memory the histogram $\, H(N)\, 
$ counting the current number of individuals sharing the same $\, N\, $, the 
number of homologous loci containing at least one copy of the ``bad'' allele, 
bit 1. This is the version where the 1-allele is dominant along the whole 
genome, exemplified in last section. An alternative version, where the 1-allele 
is recessive, is treated in the next section. At each time step, the first 
process is the killing roulette, where each individual $\, i\, $ survives 
according to a probability $\, x^{N_i+1}\, $. The number $\, x\, $ measures the 
survival probability for individuals with the wild genotype, $\, N = 0\, $. For 
the others, the survival probabilities exponentially decay with $\, N\, $. This 
is the model's selection ingredient. Note that $\, x\, $ must be {\sl strictly} 
smaller than 1, otherwise all individuals will survive forever.

	The value of $\, x\, $ is tuned in order to keep the population sample 
constant in size, and can be obtained by solving the polynomial equation

$$\sum_N H(N)\, x^{N+1} = P\, (1-b)\eqno(1)$$

\noindent {\sl before} killing anybody. (As a technical remark, we cannot use 
the solution of this equation if it surpasses some upper bound, say $\, x_{\rm 
max} = 0.999\, $: this limit simply imposes that every individual should die 
some day, no matter how good is its genetic patrimony. Extinction is the 
consequence of imposing this upper bound, as we will comment at the end of this 
section.) After the proper value for $\, x\, $ is known, we scan the 
population, $\, i = 1$, 2, 3 $\dots$ $P$. A random number $\, r_i\, $ inside 
the interval $\, (0,1)\, $ is tossed for each individual $\, i \, $: if $\, r_i 
< x^{N_i+1}\, $ it is kept in the population, otherwise it dies.

	After deaths, the next task is to include $\, (1-b)\, P\, $ newborns 
into the population. In order to construct a newborn diploid genome, first we 
toss two random parents among the survivors. (For simplicity, we do not 
consider different genders.) One parent's chromosome pair is copied and the 
following procedure is performed on the copy. An average number $\, m\, $ of 
random mutations are introduced. Each mutation acts at a random position along 
one of the two chromosomes, also taken at random. The corresponding bit is 
flipped from its current state, i.e. from 0 to 1 or from 1 to 0. The fixed 
number $\, m\, $ is not necessarily an integer, as follows. We toss a random 
number $\, M\, $ inside the interval $\, (0,2m)\, $. Then, we perform just $\, 
{\rm int}(M)\, $ mutations, where $\, {\rm int}(\dots)\, $ means the integer 
part of the argument. After that, with probability $\, {\rm frac}(M)\, $ we 
perform a last mutation, where $\, {\rm frac}(\dots)\, $ means the fractional 
part of the argument, $\, M = {\rm int}(M) + {\rm frac}(M)\, $. Also a total of 
$\, c\, $ crossings-over are performed on the diploid genome, where $\, c\, $ 
is not necessarily an integer as well: $\, {\rm int}(c)\, $ crossings are 
performed first, and a last one with probability $\, {\rm frac}(c)\, $. The 
crossing position along the diploid genome is also tossed at random. All 
results shown in the last section were obtained with $\, c = 1\, $, other 
values are treated later. After the whole process of mutations and crossings, 
we have two possible gametes. We choose one of them, also at random, to be 
passed on to the newborn. The same process is performed on the chromosome 
copies of the other parent, leading to the second newborn gamete. Then, it is 
included into the population.

\begin{figure}[!hbt]

\begin{center}
 \includegraphics[angle=-90,scale=0.5]{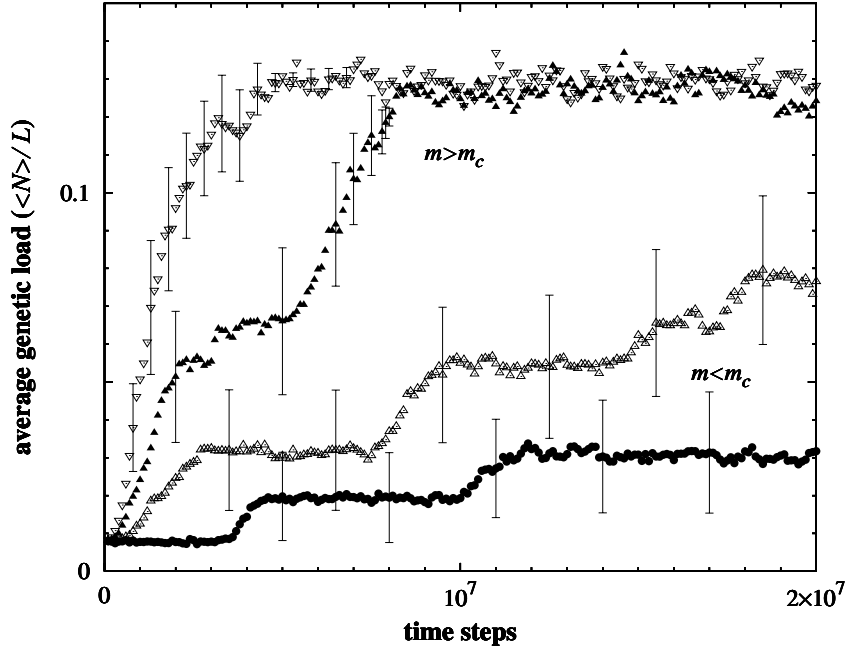}
\end{center}

\caption{Time evolution for 4 different values of $\, m\, $, all of them very 
near $\, m_c\, $. Just below the transition (two lower curves) the convergence 
is very slow. In this example, $\, c = 1.5\, $ and $\, L = 1024\, $. By 
increasing the chromosome length, the convergence of the two lower curves 
becomes still slower. However, for large $\, L\, $, the average genetic load 
$\, \langle N\rangle/L\, $ goes to zero. Contrary to that, inside the 
disordered phase one observes the two upper curves already stuck to their final 
{\sl finite} plateaux, independently of the $\, L\, $ value.}

\label{fig7}
\end{figure}

	One time step is complete after these two processes, death and birth, 
scanning the whole population. With $\, b = 0.02\, $, a complete generation 
replacement corresponds to $\, 50\, $ time steps on average. We start the 
simulation at time step $ \, t = 0\ $, with all bit-strings filled with zeroes 
(except for the hysteresis case shown later). The initial genetic distribution 
corresponds to $\, H(N=0) = P\, $ and $\, H(N \ne 0) = 0\, $. As time goes by, 
1-bits spread more or less over the population, and $\, H(N)/P\, $ eventually 
stabilises in some steady state distribution as shown in Figs.3, 4 and 5. The 
relaxation time required for stabilisation increases for increasing $\, L\, $, 
and also depends on the number $\, m\, $ of mutations performed at birth. Near 
the transition points like $\, m = 1.0439\, $ in Fig.2 the relaxation time is 
very large. The majority of our results were taken with $\, L = 1024\, $, for 
which we observed that $\, 10^7\, $ (ten million) time steps are enough, and 
decided to adopt this number as default. In some cases, specially for larger 
chromosome lengths, we have done the simulations beyond this. Due to this 
extremely slow convergence rate, each point of a plot like Fig.2 corresponds to 
approximately two entire processing days on our fastest computer processor (AMD 
Opteron 250). Recessiveness requires even more computer power.

	Fig.7 is an example of the slow convergence rate. Since the initial 
population has no 1-bits at all, the starting average genetic load $\, \langle 
N\rangle\, $ is zero. The curves show the evolution of $\, \langle N\rangle/L\, 
$ for 4 different values of $\, m\, $. The two lower curves correspond to $\, 
m\, $ below but very near the transition point $\, m_c\, $, therefore still at 
the evolutionary phase. In this case, the fluctuations are large, denoted by 
the error bars included only at some points for clarity. The two upper curves 
correspond to $\, m\, $ above but also very near the transition point $\, m_c 
\, $, therefore already at the non-evolutionary phase. Now, one gets a 
not-so-slow convergence, after which the fluctuations become smaller with error 
bars of the same size of the symbols. Fluctuations also become very small if 
one takes $\, m\, $ below but not near the transition (not shown).

	We have indentified the non-evolutionary phase on the right side of 
Fig.2 as the {\sl extinction} phase, although the population sample is kept 
with constant size. In this case, the genetic distribution among the population 
corresponds to the sharp peaks displayed in Fig.6, centred on $\, \langle 
N\rangle\, \propto\, L\, $, with a narrow relative width vanishing for large 
enough values of $\, L\, $. Therefore, the last equation (1) could be replaced 
by

$$x^{\langle N\rangle + 1} = 1 - b$$

\noindent for which the solution approaches $\, x = 1\, $ if we consider large 
$\, L\, $ and consequently large $\, \langle N\rangle\, $. However, we have 
already seen that $\, x\, $ should be {\sl strictly} smaller than 1, limited by 
some upper bound, say $\, x_{\rm max} = 0.999\, $, otherwise nobody dies. This 
upper bound is surpassed when solving equation (1) just when the run-away shown 
in Figs.5 or 6 occurs. Replacing its solution $\, x\, $ by a lower value $\, 
x_{\rm max}\, $ leads to extinction. In this way, as soon as the number $ \, 
m\, $ of mutations performed at birth surpasses the transition point $\, m_c\, 
$, not only Darwin evolution stops because all individuals become selectively 
alike, but also extinction is the next step. Therefore, within the model, we 
don't need to observe a real extinction of the population sample in order to 
identify the extinction already in course for the whole population, the genetic 
run-away or ``error catastrophe'' suffices. This approach of tuning $\, x\, $ 
in order to keep constant the population sample goes back to \cite{tb}. For 
smaller values of $\, L\, $ ($\le 64$), simulations \cite{stauffer} with fixed 
$\, x\, $ and varying $\, P\, $ give different behaviour.

\newpage
\section{Recessiveness}

	Alternatively to the 1-bit dominance, the phenotype $\, N \, $ of each 
individual can be counted as the number of loci where {\sl both} homologous 
alleles are 1-bits. This is the recessive version, much more interesting from 
the biological point of view. It allows a much larger degree of genetic 
diversity among the population, since heterozygous loci do not represent any 
handicap for the individual survival. Fig.8 is the would-be equivalent of Fig.2 
in this case.

\begin{figure}[!hbt]

\begin{center}
 \includegraphics[angle=-90,scale=0.5]{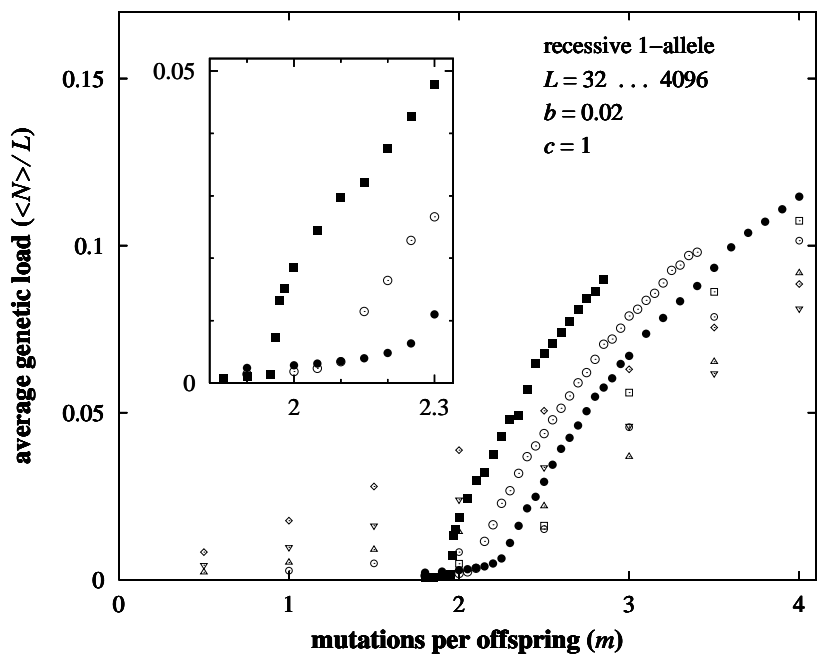}
\end{center}

\caption{Similar to Fig.2, for the recessive case, with $\, L\, $ increasing 
from bottom to top. Now, $\, N\, $ is the number of homozygous loci with both 
bad alleles. The inset blows-up the transition region for $\, L = 1024\, $ 
(full circles), 2048 (open circles) and 4096 (squares). In spite of the large 
chromosome lengths, the collapse of all curves onto a single one is not yet 
obtained. It should appear beyond $\, L = 4096\, $, defining the transition 
point (note the negative curvature which already appears for this length, when 
the full squares jump from zero to higher values, near $\, m = 2\, $).}

\label{fig8}
\end{figure}

	Larger chromosome lengths are necessary in order to observe the 
collapse of all curves onto a single one, which starts beyond $\, L = 4096\, $. 
These plots correspond to $\, 2 \times 10^7\, $ time steps and $\, P = 10000\, 
$, which were enough to obtain convergence in the case of dominant 1-bit 
allele, Figs.2 to 7. Now, it is clear that these time and population size may 
be no longer enough. For instance, from these plots, Fig.8, one could wrongly 
infer a transition point near $\, m_c \approx 2\, $. As a test, we have run 
much longer times for smaller populations $\, P = 320,\, 1000\, $ and $\, 
3200\, $, and verified the appearance of the sudden run-away already for 
smaller values $\, 1 < m_c < 2\, $, Fig.9. Some kind of staircase seems to 
appear within this interval, which is an indication that some of the 10 
independent populations considered in the averaging process have already 
undergone the run-away while others did not at the same time step. This 
behaviour also indicates the presence of hysteresis, shown in detail in next 
section. In this case, Fig.9, each point of the plot ($\, P = 3200\, $) 
comsumes more than a month of computer time, and even so we cannot be sure that 
all 10 independent populations were already genetically stabilized.

\begin{figure}[!hbt]

\begin{center}
 \includegraphics[angle=-90,scale=0.5]{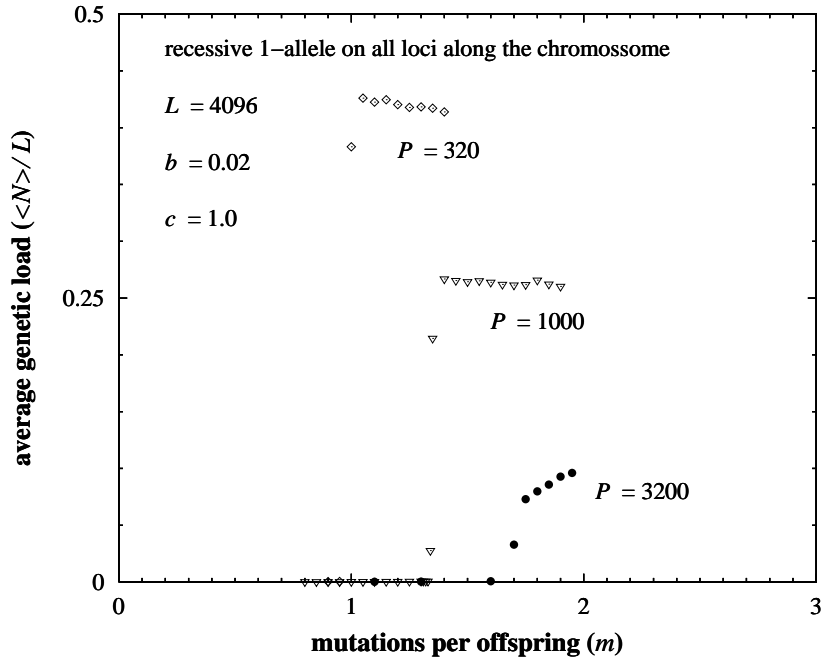}
\end{center}

\caption{Test runs with much larger times than Fig.8 (with smaller 
populations), showing the run-away (extinction phase) already appearing below 
$\, m = 2\, $.}

\label{fig9}
\end{figure}

	When, in contrast with the 1-bit allele dominant case (Fig.2), 
recessiveness is turned on (Fig.8) the current state of our simulations cannot 
precisely define the point where the phase transition occurs. However, this 
does not mean our simulations are useless for the recessive case. The phase 
transition certainly occurs in some point $\, 1 < m_c \approx 2\, $, and we can 
use the populations leading to Fig.8, for instance, in order to compare the 
features and differences between the survival and extinction phases. This is 
done in the two next sections.

\newpage
\section{Crossing-over}

	Figs.2 and 8 correspond to just one crossing-over performed during the 
gamete formation, i.e. $\, c = 1\, $. We have also tested other values.

	Fig.10 corresponds to $\, c = 0\, $, i.e. no crossing at all, for the 
case where the 1-bit allele is dominant, to be compared with Fig.2.

\begin{figure}[!hbt]

\begin{center}
 \includegraphics[angle=-90,scale=0.5]{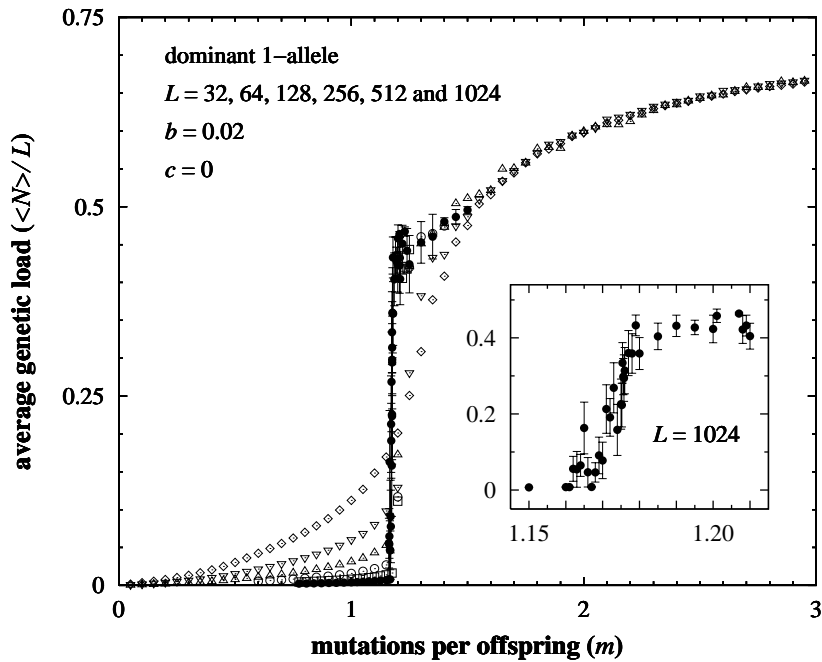}
\end{center}

\caption{First order phase transition without crossing ($\, c = 0\, $), for
dominant 1-bit alleles, to be compared with Fig.2.}

\label{fig10}
\end{figure}

	Now, the transition is clearly a first order one, with a big gap on the 
dis-order parameter at $\, m_c\, $. However, also Fig.2 seems to display a 
first order transition, but weaker as displayed in its inset there.

	Also without crossing, more interesting is the case where 1-bits are 
recessive, Fig.11. For the lower branch, filled circles, we start the whole 
process with $\, m = 1\, $ and all initial chromosomes filled with 0-bits. The 
first point on the left side corresponds to the resulting populations after $\, 
10^7\, $ time steps. Then, {\sl starting from these populations}, we tune $\, m 
= 1.05\, $ and run other further $\, 10^7\, $ time steps, getting the second 
point on the left side, and so on, increasing $\, m\, $ in steps of $ \, 0.05\, 
$. Only when we reach $\, m = 4.2\, $ the gap on the right side appears, and 
the fluctuations among the 10 independent populations become large, as denoted 
by the error bars. Soon the fluctuations shrink again at $\, m = 4.35\, $, and 
we reach the non-evolutionary phase displayed by the last four filled circles 
up to $\, m = 4.5\, $.

\begin{figure}[!hbt]

\begin{center}
 \includegraphics[angle=-90,scale=0.5]{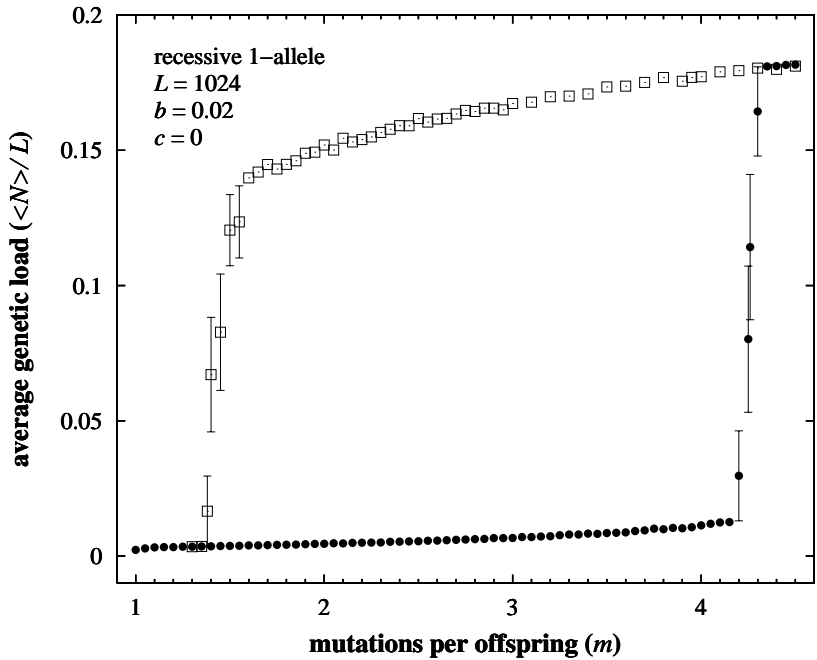}
\end{center}

\caption{Hysteresis without crossing ($\, c = 0\, $), for recessive 1-bit
alleles.}

\label{fig11}
\end{figure}

	Now, we do the reverse path, upper branch, starting from $\, m = 4.5\, 
$ again, with all initial chromosomes filled with 0-bits. After $\, 10^7\, $ 
time steps we have already reached the non-evolutionary phase, righmost open 
square (at the same position of the rightmost filled circle obtained before, 
which thus does not depend on the starting populations). Then, {\sl starting 
from these already stabilized populations}, we tune $\, m = 4.4\, $ and run 
other further $\, 10^7\, $ time steps, getting the second rightmost open square 
(also coincident with the filled circle branch), and so on, decreasing $\, m\, 
$ and running more $\, 10^7\, $ time steps for each new value. The result is 
the upper branch displayed by the open squares. Only when we reach $\, m = 
1.55\, $ this branch goes down following the gap at the left side, where the 
fluctuations (error bars) become visible again. At the end of this gap 
downwards, the upper branch meets again the lower one. In between $\, m_1 
\approx 1.4\, $ and $\, m_2 \approx 4.2\, $ the system displays a clear 
bi-stability, the equilibrium population depending on the initial one. For any 
fixed value of $\, m\, $ within this interval, the population goes to the 
evolutionary phase {\sl if} the genetic load of the initial population is small 
enough. Otherwise, it goes to the non-evolutionary phase, for the same fixed 
$\, m\, $.

	The presence of crossing-over destroys this behaviour, as shown in 
Fig.12 for $\, c = 1\, $. Now, both branches are indistinguishable, there is no 
hysteresis.

\begin{figure}[!hbt]

\begin{center}
 \includegraphics[angle=-90,scale=0.5]{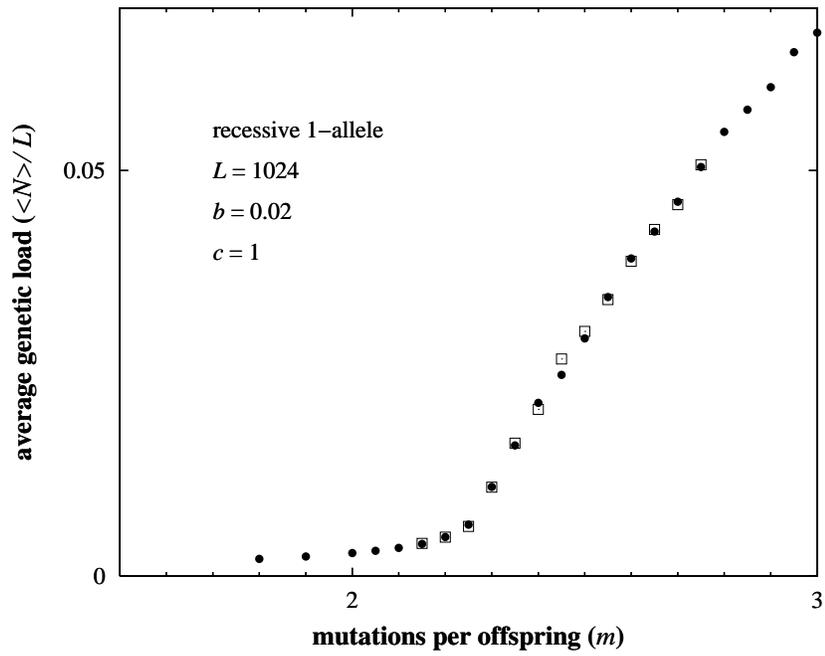}
\end{center}

\caption{Crossing destroys the hysteresis.}

\label{fig12}
\end{figure}

\newpage

	Finally, a larger number of crossings seems to have no effect, Fig.13.

\begin{figure}[!hbt]

\begin{center}
 \includegraphics[angle=-90,scale=0.5]{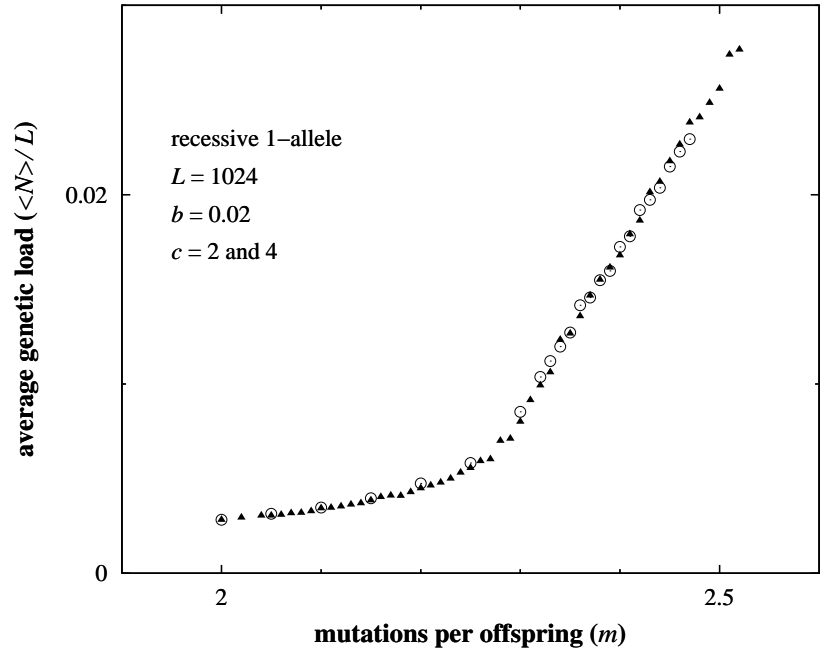}
\end{center}

\caption{The same disorder parameter obtained for more than one crossing, $\, c 
= 2\, $ (triangles) and $\, c = 4\, $ (open circles).}

\label{fig13}
\end{figure}

\newpage
\section{Heterozygosity}

        With crossings and recessive 1-bit allele, Fig.14 shows the 
heterozygosity, i.e. the fraction of heterozygous loci averaged over all 
individuals of all 10 independent populations after $\, 10^7\, $ time steps, as 
well as the corresponding fractions of both homozygous loci 11 and 00 
(homologous loci filled with the same allele 1 or 0).

\begin{figure}[!hbt]

\begin{center}
 \includegraphics[angle=-90,scale=0.5]{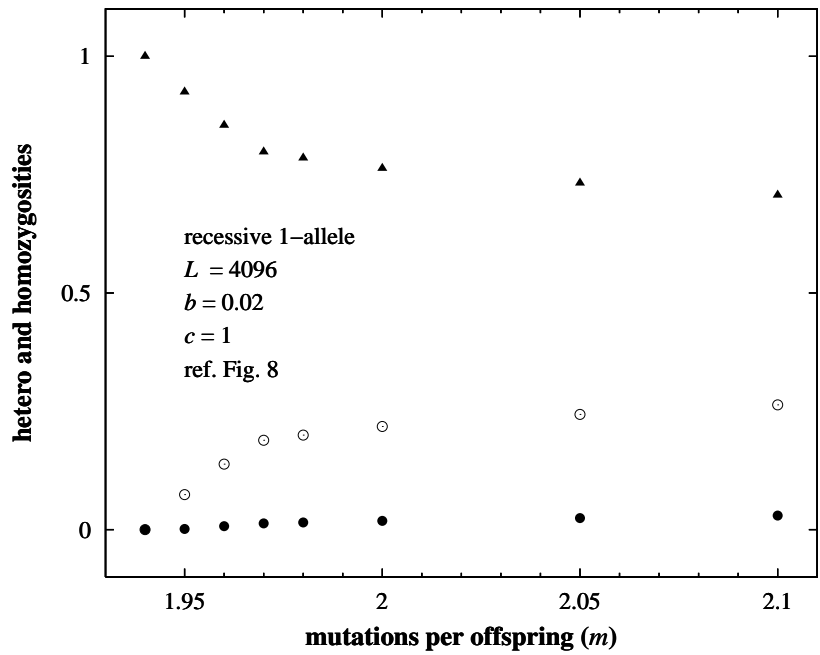}
\end{center}

\caption{Heterozygosity (open circles), homozygosity 11 (filled circles), and
homozygosity 00 (triangles), with one crossing as in Fig.8.}

\label{fig14}
\end{figure}

\newpage
        Fig.15 shows again the recessive case, now without crossing and near
the upwards jump on the right side of Fig.11. At the extinction phase, the 
heterozygosity is simply random.

\begin{figure}[!hbt]

\begin{center}
 \includegraphics[angle=-90,scale=0.5]{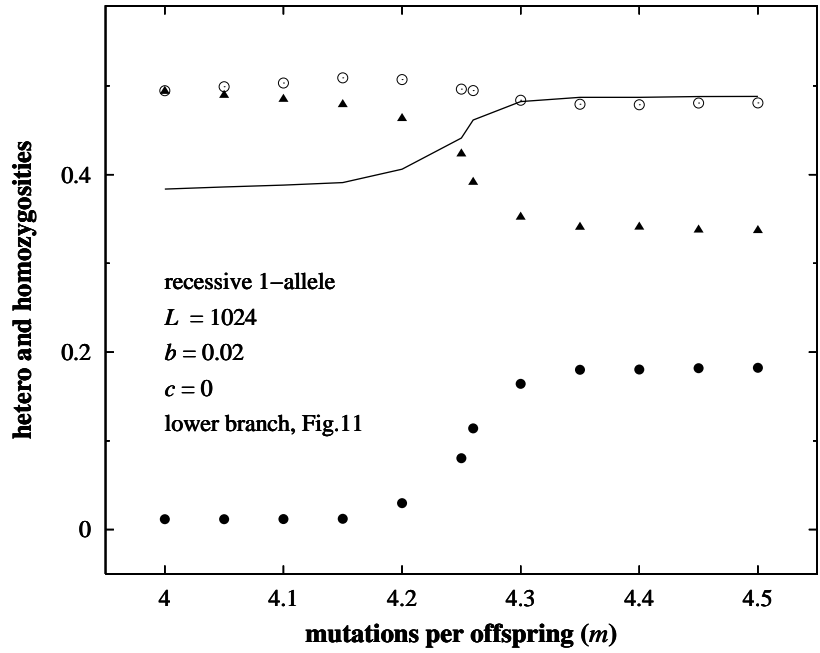}
\end{center}

\caption{Heterozygosity (open circles), homozygosity 11 (filled circles), and 
homozygosity 00 (triangles), without crossings. If the bad alleles were 
randomly distributed the heterozygosity would be given by the full line.}

\label{fig15}
\end{figure}

\newpage
        Fig.16 also refers to the case of Fig.11, now near the downwards jump 
on the left side. Again, the heterozygosity is random at the extinction phase.

\begin{figure}[!hbt]

\begin{center}
 \includegraphics[angle=-90,scale=0.5]{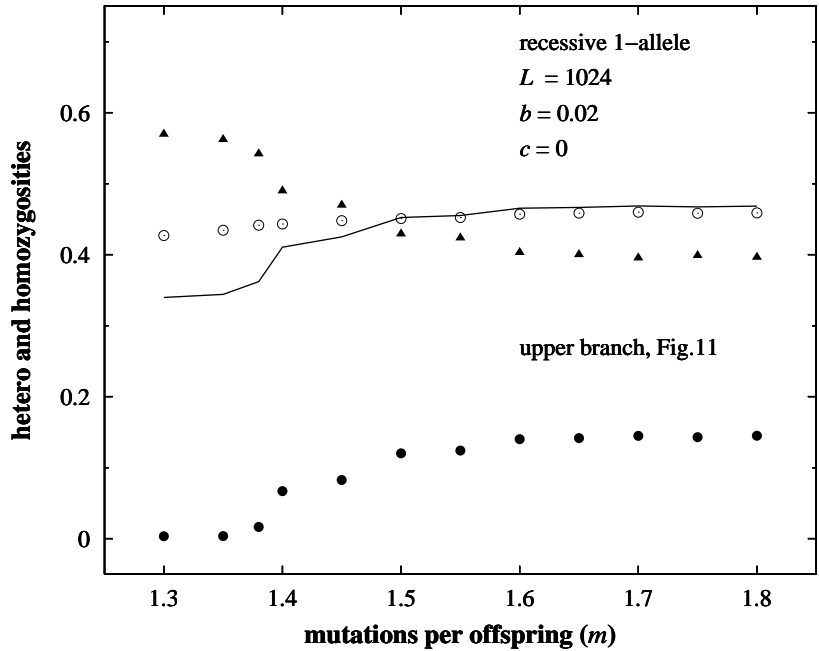}
\end{center}

\caption{Heterozygosity (open circles), homozygosity 11 (filled circles), and
homozygosity 00 (triangles), without crossings.}

\label{fig16}
\end{figure}

        Comparing Fig.14 where crossing is present to Fig.15 or 16 where $\, c 
= 0\, $, one concludes that crossing has a fundamental role: homozygosity 00 
dominates the population within the evolutionary phase, at the very left side 
of Fig.14. On the other hand, without crossing, heterozygous individuals occupy 
a large fraction of the surviving population at the evolutionary phase, left 
side of Fig.15. or 16. However, the presence of heterozygosity on the 
evolutionary phase does not mean a larger genetic diversity. On the contrary, 
without crossings, diploid individuals tend to get two complementary homologous 
chromosomes, an example of which is

$$A\hskip20pt 0\,\, 1\,\, 1\,\, 1\,\, 0\,\, 1\,\, 0\,\, 1\,\, 0\,\, 0\,\,
 0\,\, 1\,\, 0\,\, 0\,\, 1\,\, 0\,\, 1\,\, 0\,\, 0\,\, 1$$

$$B\hskip20pt 1\,\, 0\,\, 0\,\, 0\,\, 1\,\, 0\,\, 1\,\, 0\,\, 1\,\, 1\,\,
 1\,\, 0\,\, 1\,\, 1\,\, 0\,\, 1\,\, 0\,\, 1\,\, 1\,\, 0$$

\noindent where homozygous 00 and 11 loci are not shown for clarity. (Both do 
not matter for our following argument: 11 because it is anyway virtually absent 
from the evolutionary phase, according to Fig.15 or 16; and 00 because it does 
not mean any handicap.) The same kind of complementarity was also found in 
\cite{zawierta,pekalski}.

	In spite of its many ``bad'' genes, the above-exemplified individual 
has no handicap at all! Forget mutations for a while, and consider that all 
individuals become like this AB example. (This is not impossible, inbreeding 
helps.) Without crossings, their offspring are two-fold: those exactly like the 
parents (which survive), or those suffering from a strong handicap (which die). 
Surviving newborns are clonings from their parents, all of them genetically 
identical to each other. Evolution stops.

        Crossing-over, on the other hand, avoids the population to reach this 
genetic trap. Homozygosity 00 can be restored from heterozygous individuals. 
The consequence, as shown in Fig.14, is that all individuals remain genetically 
near (or at) the ``optimum'' state. Nevertheless, the population as a whole 
keeps the necessary genetic diversity to evolve: individuals with (nearly) 
complementary homologous chromosomes are exceptions, not the rule.

\newpage
\section{Conclusions}

	We have considered the evolution of sexual reproducing populations 
under a strict Darwinian-Mendelian prescription. Each individual carries a pair 
of diploid chromosomes with length $\, L\, $. Random mutations at birth are 
performed as well as crossings-over. The selection pressure removes more likely 
from the population individuals with higher numbers of harmful mutations. We 
have investigated the $\, L$-scaling properties and discovered a phase 
transition occurring at a sharply defined number $\, m_{\rm c} \approx 1\, $ of 
mutations performed at birth, the same value independent of the (large enough) 
chromosome length $\, L\, $. If the average number $\, m\, $ of mutations per 
offspring remains below $\, m_{\rm c}\, $, then the whole population survives. 
Above $\, m_{\rm c}\, $ the population undergoes a genetic meltdown, the number 
of harmful mutations explodes for all individuals (Eigen catastrophe 
\cite{eigen}), and finally the whole population is extinct. This behaviour 
comes from the dynamics of Darwin's evolution itself, under Mendel's genetic 
heritage rules, nothing more. Thus we believe it is completely general.

	The interesting point is that the average number of mutations $\, m\, $ 
performed at birth is the important parameter controlling the phase transition, 
not the mutation rate $\, m/L\, $. In reality, the DNA-copying chemical 
machinery is the same for all living beings, and works as a zipper scanning the 
whole chromosome length $\, L\, $. Therefore, apart from further 
error-correction mechanisms, the number of ``errors'' (mutations) should be 
proportional to $\, L\, $. This behaviour imposes a limit on $\, L\, $, in 
order to keep the number of mutations below the extinction transition point. 
Thus, it is not possible to evolve by simply increasing the chromosome length 
in order to store more and more genetic information, which will require an 
improvement on repllication fidelity. Moreover, considering only the coding 
parts of our genetic information, the real number of mutations per genome is 
indeed near $\, m_{\rm c} \approx 1\, $, in agreement with our results.

	Also interesting is the absence of such a transition for haploid, 
asexual reproducing populations \cite{jpc}. In this case, the same genetic 
meltdown also occurs, but it can be circumvented by artificially increasing the 
population proportionally to $\, L^\alpha\, $, with $\, \alpha \approx 2.3\, $ 
\cite{jpc}. For the present case of sexual reproducing populations, the 
transition remains no matter how large are the populations we have tested.

\end{document}